\documentclass[10pt]{iopart}

\usepackage{graphicx}
\usepackage{dcolumn}
\usepackage{bm}
\usepackage{mathrsfs}
\usepackage{amssymb}
\usepackage[breaklinks=true,colorlinks=true,linkcolor=blue,urlcolor=blue,citecolor=blue]{hyperref}

\begin{document}

\title[Spontaneous parametric down conversion in chirped, aperiodically-poled crystals]
{Spontaneous parametric down conversion in chirped, aperiodically-poled crystals}

\author{Xochitl S\'anchez-Lozano, Jose Luis Lucio}

\address{Departamento de F\'isica-DCI,
Universidad de Guanajuato, P.O. Box E-143, 37150, Le\'on, Gto.,
M\'exico.}

\ead{xochitl@fisica.ugto.mx}

\begin{abstract}
We present a theoretical analysis of the process of spontaneous parametric down conversion (SPDC) in a non-linear crystal characterized by a linearly-chirped $\chi^{(2)}$ grating along the direction of propagation.    Our analysis leads to an expression for the joint spectral amplitude, based on which we can derive various spectral-temporal properties of the photon pairs and of the heralded single photons obtained from the photon pairs, including: the single-photon spectrum, the chronocyclic Wigner function and the Schmidt number.  The simulations that we present are for the specific case of a collinear SPDC source based on a PPLN crystal with the signal and idler photons emitted close to the telecom window.  We discuss the mechanism for spectral broadening due to the presence of a linearly chirped $\chi^{(2)}$ grating, showing that not only the width but also to some extent the shape of the SPDC spectrum maybe controlled.  Also, we discuss how the fact that the different spectral components are emitted on different planes in the crystal leads to single-photon chirp.
\end{abstract}

\pacs{42.50.-p, 3.65.Ud, 42.50.Dv, 3.65.-w, 3.67.-a}
\vspace{2pc}
\noindent{\it Keywords}: spontaneous parametric downonversion, entanglement, periodic poling
\ioptwocol

\section{Introduction}

Efficient and controllable generation of non-classical light, in particular photon pairs, is essential for the implementation of various quantum information processing schemes. In this regard, spontaneous parametric down-conversion (SPDC) is routinely employed, a process in which single photons from a pump beam are annihilated in a second-order ($\chi^{(2)}$) nonlinear crystal leading to the generation of signal and idler photon pairs. This process leads to quite a remarkable flexibility in the scope of spatial and spectral properties which may characterize the photon pairs. Indeed, careful design of the attributes of the nonlinear crystal and of the pump beam can result in photon states with engineered properties for specific needs.

In this paper, we will focus on the use of quasi-phase-matched (QPM) crystals, also referred to as periodically poled crystals, i.e. those in which the three waves involved in the SPDC process (pump, signal, and idler), experience a grating in the $\chi^{(2)}$ nonlinearity along the direction of propagation. This grating is such that $\chi^{(2)}$ oscillates between positive and negative values with a characteristic poling period $\Lambda$.   In particular, in this paper, we will restrict attention to a collinear geometry in which these three waves co-propagate along a given axis of the crystal. Periodic poling leads to a number of key advantages over the more traditional birefringent phasematching, for which the three waves involve two orthogonal polarizations and where matching of the phases gained in propagation through the crystal is attained through careful specification of the direction of propagation in the anisotropic crystal. Advantages of QPM include: i) the possibility of accessing larger diagonal elements of the $\chi^{(2)}$ tensor, which can lead to a larger flux  (which is proportional to the square of the effective $\chi^{(2)}$) and also leads to co-polarized pump, signal and idler waves, and ii) because the direction of propagation must no longer be specified in a specific manner to attain phasematching, it becomes possible to choose a direction of propagation so as to suppress Poyinting vector walk-off; note that this permits the use of longer crystals compared with birefringent phasematching.

In this work we will in particular study an extension of quasi-phase-matching, involving a slight aperiodicity in the $\chi^{(2)}$ grating, so that the poling period has a variation across the crystal in the form of a linear chirp. As will be discussed below, the poling period is linked to the phasematched signal and idler frequencies which may be generated. The availability of a \emph{range} of periods in a chirped $\chi^{(2)}$ grating thus leads to multiple phasematched signal and idler frequencies, and thus to the emission of broadband photon pairs~\cite{Giuseppe2002,Torres2004, Odonnell2007}. There are other possible avenues for the generation of particularly broadband SPDC light.  One possibility is the use of a very short crystal, for which phasematching constraints are relaxed. However, a very short crystal also implies a prohibitively low flux~\cite{Dauler1999}. Another possibility is the use of a specific crystal, as well as specific pump and SPDC frequencies so that certain conditions on the crystal dispersion are met which promote a large bandwidth~\cite{Odonnell2007}. While this method is highly effective, unfortunately the specific conditions needed on the crystal and on the pump/SPDC frequencies in practice makes this avenue impractical. In contrast, the use of a chirped quasi phase matched crystal can be designed to yield a large emission bandwidth for arbitrary poled crystals and for arbitrary pump/SPDC frequencies (assuming of course that the required $\chi^{(2)}$ grating can be fabricated).

Crystals involving a chirped $\chi^{(2)}$ grating in the spatial coor\-dinate along the direction of propagation have also been considered in previous works~\cite{Carrasco2004, Harris2007, Nasr2008}. Theoretical results have been reported, using diverse assumptions about the pump beam or different approa\-ches to perform the calculation. Thus, for exam\-ple, Fraine~\cite{Fraine2012} considers an incident monochromatic plane wave and his treatment is restricted to the production of collinear signal and idler modes. The author assumes that the spectrum produced by the chirped crystal can be approximated by the spectrum arising from a collection of independent, uniform, non linear crystals. In this way the SPDC-type II spectrum depending on the crystal chirp and source localization is obtained. In~\cite{Antonosyan2012} a similar treatment is applied to perform a detailed description of SPDC process with the implementation of discrete Gauss sums to quantum system.

The state of SPDC photon pairs can be engineered through the attributes of the crystal, as discussed in the previous paragraph, or through the attributes of the pump.  While the focus of this paper is the use of a micro-structured crystal, the use of a structured pump, spatially as in ~\cite{Ramirez03,Cruz02}, or temporally as in~\cite{Sanchez2012}, also leads to the possibility of tailoring the SPDC two-photon state. The current paper may be regarded as an extension of our earlier papers~\cite{Sanchez2012,Yasser07} where the use of a \emph{chirped pump} (through the action of group velocity dispersion) is used rather than a \emph{chirped crystal}.

\section{Spontaneous parametric down conversion in chirped quasi-phasematched nonlinear crystals}

The most widely used method for generating biphotons is through the SPDC process, which occurs when an optical crystal with a second-order non-linearity is illuminated by a laser beam, referred to as the pump. Individual photons from the pump beam may then be annihilated, leading to the emission of two lower-frequency photons, where the two photons in a given pair are typically referred to as signal and idler. These photon pairs may be entangled in any of the photonic degrees of freedom, including polarization, time-frequency and transverse position-momentum. In this paper we concentrate on the spectral degree of freedom; specifically, we assume that appropriate spatial filtering on the signal and idler modes is used so that only specific directions of propagation are retained.

The two-photon state for the SPDC process following a standard perturbative approach and assuming specific directions of propagation, i.e. spatially-filtered so that only those $k$ vectors which are parallel to the pump direction of propagation are retained, can be expressed as

\begin{equation}
\label{State} |\Psi\rangle = |\mbox{vac}\rangle+\eta\int d\omega_s \int d\omega_i f(\omega_s, \omega_i)|\omega_s\rangle|\omega_i\rangle,
\end{equation}

\noindent where $\eta$ is a constant related to the conversion efficiency, $|\omega_\mu\rangle = \hat{a}^\dag_\mu(\omega)|0\rangle$ with $\mu=i,s$ and $|0\rangle$ represents the va\-cuum state. The joint spectral amplitude (JSA) $f(\omega_s, \omega_i)$  is written as the product of two distinct quantities

\begin{equation}
\label{JSI} f(\omega_s, \omega_i) = \Phi(\omega_s, \omega_i)\gamma(\omega_s + \omega_i),
\end{equation}

\noindent where $\Phi(\omega_s, \omega_i)$ is the phase mat\-ching function (PMF) and $\gamma(\omega_s + \omega_i)$ is the pump envelope function (PEF). Note that while for a monochromatic pump, the PEF function in Eq.~\ref{JSI} is becomes a delta function centered at the pump frequency, an ultrafast pump (as will be assumed in simulations below) corresponds to the coherent addition of different pump spectral components over a significant bandwidth. We model the PEF as a Gaussian function with bandwidth $\sigma$: $\gamma(\omega_s + \omega_i) = \textrm{exp}[-(\nu_s+ \nu_i)^2/\sigma^2]$, where $\nu_\mu=\omega_\mu-\omega_c$ are the frequency detunings with $\omega_c$ the central generation frequency.
If the joint amplitude is normalized so that $\int d\omega_s d \omega_i |f(\omega_s,\omega_i)|^2 =1$,  $|f(\omega_s,\omega_i)|^2$ represents a joint probability distribution for the emission of photon pairs with frequencies $\omega_s$ and $\omega_i$ and is known as the joint spectral intensity (JSI).

The freedom to tailor the two-photon states arises from the crystal and pump properties, which enter the calculation through the JSA. In order to engineer the state produced by SPDC we rely upon the PMF that incorporates properties of the crystal and the pump field

\begin{eqnarray}
\Phi(\vec{k}_s, \vec{k}_i) = \int dV \textrm{d}(\vec{r})\tilde{\alpha}(\vec{r},\omega_s + \omega_i)  e^{-i(\vec{k}_s+ \vec{k}_i)\cdot \vec{r}}.\label{pm}
\end{eqnarray}

\noindent where $\tilde{\alpha}(\vec{r},\omega_p)$ is the spatial distribution of the pump, which we will later assume to be in the form of a Gaussian beam and where $\textrm{d}(\vec{r})$ represents nonlinearity of the crystal, which in our analysis will have a spatial dependence associated with the $\chi^{(2)}$ grating.

An important challenge in the utilization of the SPDC process is that the conversion efficiency tends to be low and thus the resulting photon-pair flux tends to be limited. The SPDC flux is proportional to the crystal length and to the pump power. Increasing the crystal length $L$ so as to boost the SPDC flux can be problematic, since there is also an inverse relationship between $L$ and the SPDC spectral bandwidth.  An alternative approach is the use of an aperiodic quasi phasematching (QPM) crystal - in particular one with a linear chirp, i.e. involving a linear variation of the QPM period  along the axis of propagation. Chirped quasi phasemathcing (CQPM) allows, as will be studied below, the fulfilment of the phasematching condition in a wider spectral range \cite{Nasr2008}.

For an unchirped QPM crystal, the nonlinearity alternates in sign over subsequent crystal segments of length $\Lambda/2$, where $\Lambda$ is the QPM period. One may then express the nonlinearity as a function of the propagation distance $\textrm{d}(z)$ as a Fourier series~\cite{Clausen1999,Torres2004}, where each term in the series has the form

\begin{eqnarray}
\textrm{d}(z) \propto \frac{2}{\pi m} e^{-i m K_0 z},\label{varsin}
\end{eqnarray}

\noindent where $K_0 = 2\pi/\Lambda$ and where $z=0$ corresponds to the second face of the crystal. Each term in the Fourier series leads to a phasematching condition involving the refractive index $n_p / n_s / n_i $ and the frequency  $\omega_p$/$\omega_s$/$\omega_i$ (with $\omega_i=\omega_p-\omega_s$) for the pump/signal/idler, as follows

\begin{equation}
\frac{n_p \omega_p}{c}-\frac{n_s \omega_s}{c}-\frac{n_i \omega_i}{c}+m K_0 = 0,
\end{equation}

\noindent where $c$ is the speed of light in vacuum.  If this condition is fulfilled for a given value of $m$, which is then referred to as $m$th order QPM, this in turn leads to a given phasematched combination of pump and generation frequencies. Note that if the phasematching condition is fulfilled for more than one value of $m$, the associated combinations of pump and generation frequencies tend to be distinct, so that in a given phasematched experimental situation, a single value of $m$ tends to be relevant.

Note also that because the argument of the exponential in Eq.~\ref{varsin} is proportional to $m / \Lambda$, $m$th-order QPM behaves as $1$st order with a shorter period $\Lambda/m$. Therefore, higher-order QPM presents an alternative to QPM gratings with very short periods, which may be difficult to fabricate. However, because the effective nonlinearity is proportional to $1/m$, progressively higher QPM orders also lead to progressively lower conversion efficiencies.

A linear chirp in the QPM grating can then be introduced by substituting the fixed grating spatial frequency $K_0$ by the function $K(z)=K_0+D(z_0+z)$ where $K_0=2 \pi / \Lambda_c$, $D$ represents the chirp parameter, and $z=-z_0$ represents the spatial coordinate for which $K(z)=K_0$ (i.e. for which the local QPM period is $\Lambda_c$); $z_0$ will be referred to as the reference $z$ coordinate.

Our crystal of length $L$ covers the region of the chirped grating, along the propagation axis, from $z=A$ to $z=A+L$, so that the value of $z_0$ may in fact lie inside or outside of the crystal. The nonlinearity as a function of $z$, for $m=1$, may then be written as

\begin{eqnarray}
\textrm{d}(z) \propto e^{-i[K_0 + D(z_c + z)]z},\label{varsin+chirp}
\end{eqnarray}

Note that in this case, the effective, local period may be considered to be a function of the propagation distance according to  $\Lambda(z)= 2\pi/[K_0 + D(z_0 + z)]$, which provides a collection of phase matching conditions over the crystal length. The existence of multiple phasematching conditions can lead to broadband biphoton generation \cite{Nasr2008}. The chirped poling pattern could be chosen with a special phase relation among the various spectral components thereby allowing the biphoton wavepacket to be compressed using the techniques of ultrafast optics \cite{Sensarn2010}. The generation of such ultrabroadband nonclassical light opens the door for the production of a high flux of biphotons with spectrally non-overlapping spectral components which could find application, for example, in implementations of  optical coherence tomography where the resolution is inversely proportional to the bandwidth used.

Note that if all three waves are co-polarized, we may access the diagonal elements of the $\chi^{(2)}$ tensor, which tend to have higher values than the off-diagonal terms resulting in higher emission rates. Therefore, in our analysis we assume type 0 phasematching (i.e. co-polarized pump, signal, and idler waves). We also assume a collinear geometry in which the three waves co-propagate along a single axis. For a pump in the form of a Gaussian beam and a nonlinearity vs propagation distance given according to  Eq.~\ref{varsin+chirp}, we obtain

\begin{eqnarray}
\nonumber \Phi(\omega_s,\omega_i) &=& \sqrt{\frac{\pi}{2 D}}e^{[-i\frac{(\Delta k- Dz_0)^2}{4D}} \\ \nonumber
&\times& (\mathrm{erf}[\frac{(-1)^\frac{1}{4}(2BD - \Delta k + Dz_0)}{2\sqrt{D}}] \\
&-& \mathrm{erf}[\frac{(-1)^\frac{1}{4}(2AD-\Delta k + Dz_0)}{2\sqrt{D}}]),\label{pmfqpm+chirpcolineal}
\end{eqnarray}

\noindent where $z=A$ and $z=B$ represent the boundaries of the crystal, so that  $L=B-A$;  $\Delta k= k_p(\omega_s+\omega_i) -k_s(\omega_s)-k_i(\omega_i) -2\pi/\Lambda_c$ is the phase mismatch, and the error function is defined by $\mathrm{erf}(x)\equiv\frac{2}{\sqrt{\pi}}\int_{0}^{x} e^{-t^2} dt$. In the limit $D\rightarrow 0$ we obtain the result for an unchirped QPM crystal. Furthermore, with the necessary simplifications, this expression agrees with that calculated through other methods by Harris in \cite{Harris2007}.

From this point onward, we use the expression which results from the specific crystal boundaries $A=-L$, $B=0$ and from expressing the reference coordinate as $z_0=rL$, so that Eq. (\ref{pmfqpm+chirpcolineal}) simplifies to

\begin{eqnarray}
\nonumber \Phi(\omega_s,\omega_i) &\sim& \sqrt{\frac{\pi}{2\xi}}e^{i\frac{(-x+ r\xi)^2}{4\xi}} (\mathrm{erf}[\frac{(-1)^\frac{1}{4}(-x +r \xi)}{2\sqrt{\xi}}] \\
&-& \mathrm{erf}[\frac{(-1)^\frac{1}{4}(-x +\xi(r-2))}{2\sqrt{\xi}}]),\label{pmfqpm+chirpcolineal2}
\end{eqnarray}

\noindent in terms of the parameters: i)  $x=L\Delta k_p$, which corresponds to an dimensional phase mismatch  and ii) $\xi=DL^2$ which corresponds to an dimensional effective chirp parameter.

Based on Eq.~\ref{pmfqpm+chirpcolineal2}, we have performed simulations, in which we vary the chirp parameter $D$ and the reference coordinate $z_0$ for a fixed crystal length $L$.  We investigate the effects of varying these parameters on the the single-photon spectrum,  on the chronocyclic Wigner function, and on the Schmidt number.

\section{Results: Joint spectral intensity and single-photon spectrum}

\subsection{Selecting the crystal and pump properties}

There are a number of aspects that should be taken into account when selecting the properties of the nonlinear crystal and the pump beam.

Note that the crystal temperature modifies the two-photon state in two ways. On the one hand, the refractive index depends on the temperature, and therefore the dispersion experienced by each of the thee waves likewise depends on the temperature. Because phasematching conditions depend critically on crystal dispersion, temperature will influence the resulting phasematching properties. On the other hand thermal expansion/contraction leads to a temperature dependence of the crystal dimensions, so that a periodically poled crystal will exhibit a period which depends on the temperature. This means that phasematched frequencies will tend to shift with temperature changes; in practice, it is often necessary to control the crystal temperature so as to control the SPDC emission frequencies. For our analysis, we have assumed a fixed temperature of 25 degrees Celsius.

The PMF, see Eq. (\ref{pmfqpm+chirpcolineal2}), depends on the following parameters: $x=L\Delta k_p$, $\xi=DL^2$, and $r$. Note that $\xi$ scales linearly with the $D$ and quadratically with $L$ so that both of these two quantities, together, determine the effective chirp.  It is helpful, experimentally, to have two different parameters which determine the overall effect of crystal chirp.

The QPM period is limited by current poling technology; for example in the case of PPLN it is difficult to achieve periods shorter than $5~\mu$m  \cite{Loza2001, Cudney2002}.  Also, if the pump pulse spatial width $\delta z$ (linked to its temporal width $\delta t$ as $\delta z= c \delta t$) is short enough, when compared to the crystal length $L$, so that the pulse scans the variation of the nonlinearity across the crystal as it propagates, then additional temporal shaping of the bi-photon is expected. Because the local period experienced by the pump pulse as it scans the crystal determines the emission frequency, this can lead to temporal-spectral correlations, i.e. to chirp, in the biphoton. In other words, for sufficiently short pump pulses, chirp in the QPM crystal can translate into single-photon chirp.

In addition to the crystal length $L$, the period $\Lambda_c$ and the chirp parameter $D$, the $z$ coordinate reference $z_0$ will impact the resulting photon-pair properties. In our simulations presented below, we show the effect of varying $z_0$.

For the numerical analysis we have assumed a periodically poled lithium niobate crystal of length $L=5$ mm with the chirp parameter ran\-ging from $D = 3\times 10^{-7}~\mu \textrm{m}^{-2}$ to $D = 7\times 10^{-6}~\mu \textrm{m}^{-2}$. The frequency dependence of the refractive index is introduced through the Sellmeier equation for the extraordinary ray, taken from Ref.~\cite{Jundt1997}, which is valid in the $\lambda=0.4 - 5~\mu$m  range.

We have assumed that the pump beam is centered at $800$ nm with a full width at half maximum (FWHM) of 10 nm (this could be obtained from a mode-locked Titanium Sapphire laser); we have assumed a central poling period given by $\Lambda_c= 20.33~\mu$m. In our simulations, the frequency-degenerate photon pairs are generated collinearly with the pump beam, with all three waves polarized so that they constitute extraordinary rays (type 0 phasematching). This configuration produces anti-correlated biphotons at $1600$ nm, near the telecom window.

\subsection{The joint spectral intensity and the single photon spectrum}

\begin{figure}[h!]
\begin{center}
\centering\includegraphics[height=4.5cm, width=5.5cm]{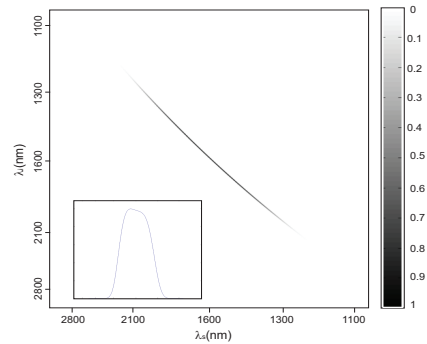}
\end{center}
\par
\caption{Joint spectrum $|f(\omega_s,\omega_i)|^2$ for SPDC degenerate gene\-ration at 1600 nm, using a 5 mm PPLN crystal pumping at 800 nm. Inset shows the corresponding single spectrum.} \label{1}
\end{figure}

In this section we report numerical simulations for: i) the joint spectral intensity (JSI), or $|f(\omega_s,\omega_i)|^2$, obtained from Eq. (\ref{JSI}), and ii) the single photon spectrum (SPS), which corresponds to the marginal distribution obtained by integrating $|f(\omega_s,\omega_i)|^2$ over $\omega_i$. In order to show the effects of QPM chirp, we focus on the second of these quantities. In our numerical simulations, no approximations are required so that our results include dispersion effects to all orders, in particular group velocity and group velocity dispersion effects. Positive and negative chirp values as well $z_0$ values corresponding to locations both inside and outside the crystal are considered. Figure \ref{1} corresponds to the spectrally anti-correlated state generated in the absence of chirp, i.e. for an unchirped PPLN crystal. The inset shows the single photon spectrum with an emission bandwidth of $\sim0.60~\mu$m. In this case, the phasematching condition for the gene\-ration of biphotons centered at $1600$ nm is satisfied on any given plane within the crystal. This is no longer true for $D \neq 0$, for which different emission frequencies will be emitted on different planes within the crystal.

\bigskip

\begin{figure}[h!]
\begin{center}
\centering\includegraphics[height=4.6cm, width=7.0cm]{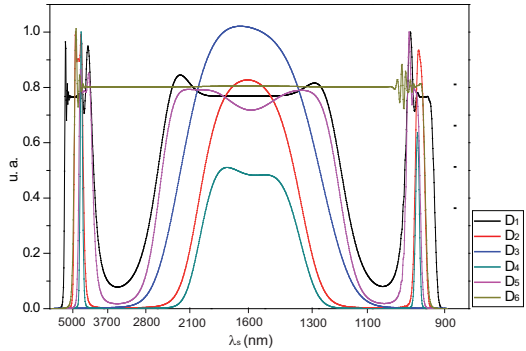}
\end{center}
\par
\caption{Single photon spectrum for $z_0=0.5 L$ and different values of chirp: black line, $D_1=-7\times10^{-6} ~\mu \textrm{m}^{-2}$; red line, $D_2=-2\times10^{-6} ~\mu \textrm{m}^{-2}$; blue line, $D_3=-3\times10^{-7} ~\mu \textrm{m}^{-2}$; cyan line, $D_4=3\times10^{-7} ~\mu \textrm{m}^{-2}$; pink line, $D_5=2\times10^{-6} ~\mu \textrm{m}^{-2}$ and green line, $D_6=7\times10^{-6} ~\mu \textrm{m}^{-2}$.} \label{2}
\end{figure}

Next, consider a crystal with non-vanishing chirp. Our results indicate that an important effect of crystal chirp is an increase, which can be very substantial, of the emission bandwidth. This increased emission bandwidth ($\Delta \omega$) becomes apparent when plotting the SPS.   Figure \ref{2} shows the SPS for $z_0 = 0.5 L$ (which corresponds to a local QPM period $\Lambda = \Lambda_c$, or equivalently to $K(z)=K_0$ at the center of the crystal), and for different values of the chirp parameter ($D_{6,1}=\pm 7\times10^{-6} ~\mu \textrm{m}^{-2}$, $D_{5,2}=\pm 2\times10^{-6} ~\mu \textrm{m}^{-2}$, $D_{4,3}=\pm 3\times10^{-7} ~\mu \textrm{m}^{-2}$). The fo\-llo\-wing characteristics are worth remarking: {\it i}) the SPS shows a central structure and two side peaks, {\it ii}) for positive chirp the maximum bandwidth is $\Delta\omega_{D_6}\approx 1.62~\mu$m larger than the emission bandwidth in the absence of chirp, while for negative chirp the broade\-ning is smaller. {\it iii}) depending on the  chirp value, the probability distribution shows a dip at the central wavelength of generation, in this case at 1.6$~\mu$m.

\begin{figure}[h!]
\begin{center}
\centering\includegraphics[height=8.5cm, width=7cm]{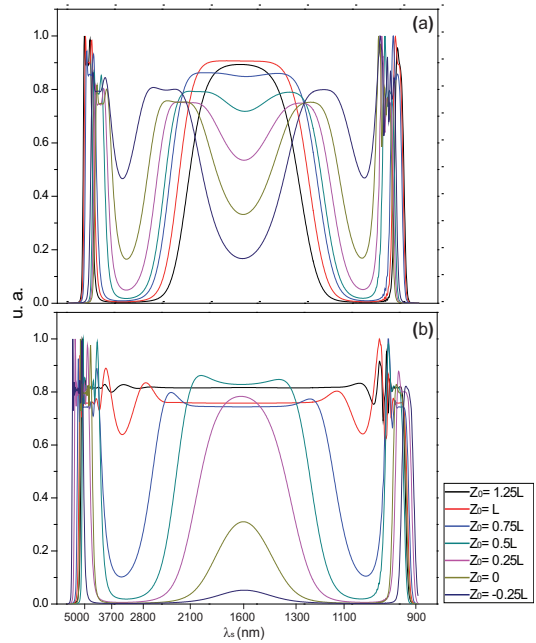}
\end{center}
\par
\caption{Single photon spectrum for different longitudinal positions of the central period: black line, $r=1.25$; red line, $r=1$; blue line, $r=0.75$; cyan line, $r=0.5$; pink line, $r=0.25$, and green line, $r=0$; and navy line, $r=-0.25$. Panel (a) with $D=2\times 10^{-6} ~\mu \textrm{m}^{-2}$, and panel (b) with $D=-5\times 10^{-6} ~\mu \textrm{m}^{-2}$.} \label{3}
\end{figure}

In order to put in evidence the relevance of the parameter $z_0$, in Figure \ref{3} we report the SPS for fixed chirp value and $r=1.25, 1.0, 0.75$, $0.5, 0.25, 0$, and $-0.25$. Panel (a) shows the results when $D=2\times10^{-6} ~\mu \textrm{m}^{-2}$ and panel (b) $D=-5\times10^{-6} ~\mu \textrm{m}^{-2}$. In this case the following points can be remarked: {\it i}) large bandwidths can be produced with both signs of chirp,  ii) when $z_0$ is outside the crystal, but close to the face where the pump laser impinges, a larger bandwidth is generated when using a negative chirp. This can be seen by comparing panels (a) and (b) of Figure \ref{3}. The absolute value of the chirp parameter is larger in panel $(b)$ than in panel $(a)$; these values were selected in this way, since a larger $|D|$ is needed for $D<0$, as compared to the $D>0$ case in order to yield a similar degree of broadening, {\it iii}) if $z_0$ is located inside and close to the far side of the crystal, a redistribution of the emission frequencies occurs so that generation of the central wavelength is comparatively suppressed while frequencies far from the degenerate frequency are produced.

\subsection{CWF and frequency entanglement}

\noindent The chronocyclic Wigner function (CWF) provides spectral and temporal information about single photons (say in the signal mode) heralded by the detection of a single photon in the conjugate mode. The CWF has a close connection with the signal-mode reduced density matrix $\hat{\rho}_s$  and hence the purity $\mbox{Tr}(\hat{\rho}_s^2)$ \cite{Sanchez2012}. The CWF is given as follows

\begin{eqnarray}
W_s(\omega,t)&=\int d\omega_0  \int d \omega' f\left(\omega_0,\omega+\frac{\omega'}{2} \right)  \nonumber \\ &\times
f^*\left(\omega_0,\omega-\frac{\omega'}{2} \right) e^{i \omega' t}
\label{Eq:CWF}
\end{eqnarray}

The single-photon spectrum can be obtained by integrating the CWF over time, and is equivalent to the result obtained by the marginal distribution of the JSI, as used earlier in this paper. The single-photon temporal intensity can be obtained by integrating the CWF over frequency, and is equivalent to the marginal distribution of the joint temporal intensity.

\begin{figure}[h!]
\begin{center}
\centering\includegraphics[height=7cm, width=6.5cm]{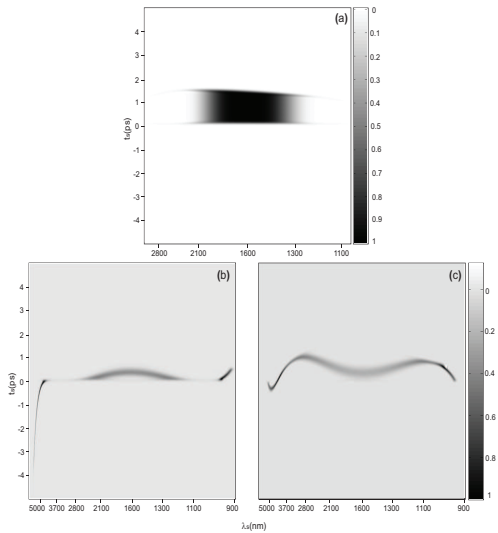}
\end{center}
\par
\caption{Single-photon chronocyclic Wigner function with $z_0 = 0.5$L. Panels (a) $D=0$, (b) $D=-5\times10^{-6} ~\mu \textrm{m}^{-2}$, and (c) $D=3\times10^{-6} ~\mu \textrm{m}^{-2}$.} \label{4}
\end{figure}

Figure \ref{4} shows the CWF, which characterizes the spectral and temporal properties of heralded single-photon wavepackets in the signal mode, as heralded by the detection of a conjugate idler photon; this plot assumes  $z_0 = 0.5 L$ (for other values of $z_0$  similar results are obtained) and we consider three different  values of the chirp parameter $D$. Panel (a) corresponds to $D=0$; panel (b) corresponds to $D=-5\times10^{-6} ~\mu \textrm{m}^{-2}$; panel (c) corresponds to $D=3\times10^{-6} ~\mu \textrm{m}^{-2}$. Note that the spectral broadening of the single-photon wavepacket, with respect to the unchirped $D=0$ case, becomes evident in these plots. Also, note from these plots that while in the $D=0$ case, basically all available SPDC frequencies are emitted at any given instant during the duration of the single-photon wavepacket, for the $D \neq 0$ cases, specific frequencies are emitted at specific times, i.e. the single photon wavepacket becomes chirped in a specific manner.  It becomes clear from panels b) and c) that the sign and magnitude of $D$ controls the specific spectral-temporal correlations which appear in the single-photon wavepacket.

\begin{figure}[h!]
\begin{center}
\centering\includegraphics[height=5cm, width=6.5cm]{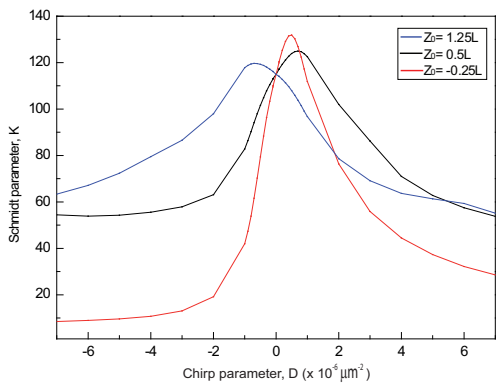}
\end{center}
\par
\caption{Schmidt parameter vs. chirp in the chirp parameter $D$ for different $z_0$ values.  Red line, $z_0=-0.25$ L, black line $z_0=0.5$ L, and blue line $z_0= 1.25$ L.} \label{5}
\end{figure}

Entanglement quantification is a central element in the charac\-teri\-zation of the quantum state produced. The relationship between photon-pair entanglement and the  single-photon purity is well known. The Schmidt number $K$ quantifies the degree of photon-pair entanglement while the single-photon purity is $\mbox{Tr}(\hat{\rho}_s^2)$ is in fact the reciprocal of $K$. Thus, for a factorable state characterized by $K = 1$, heralded single photons are pure, i.e.  $\mbox{Tr}(\hat{\rho}_s^2)=1$. Meanwhile, for $K \rightarrow\infty$ i.e for a highly entangled photon pair, the heralded single photon becomes highly impure, i.e.  $\mbox{Tr}(\hat{\rho}_s^2) \rightarrow 0$ \cite{Uren2005}. We have computed the single photon purity \cite{Sanchez2012} produced in the chirped PPLN crystal with chirp values $D_a=0$, $D_b=-5\times10^{-6} \mu \textrm{m}^{-2}$; and $D_c=3\times10^{-6} \mu\textrm{m}^{-2}$  with the \-following\- results $\textrm{p}_{D_a}=0.0087, \textrm{p}_{D_b}=0.0184$, and $\textrm{p}_{D_c}=0.0116$. We have computed $K$ for different source configurations, as shown in Fig~\ref{5}~\cite{Sanchez2012, Law2000}.  Note that while in the regimes considered here the two-photon states are far from being factorable, $K$  exhibits a dependence on the crystal chirp parameters $D$ and $z_0$ and thus crystal chirp could be used as an effective tool to vary the degree of entanglement.

\section{Conclusions}

The use of periodically poled, or quasi-phasematched, crystals constitutes an important alternative to birefringent phase mismatching which can lead to particularly bright SPDC sources. The use of an aperiodicity in the poling, in the form of a chirped $\chi^{(2)}$ grating leads to the fulfilment of multiple phasematching conditions over a range of frequencies, which in turn leads to the emission of particularly broadband SPDC photon pairs. In this paper we present a theoretical analysis of SPDC based on chirped periodically-poled crystals. We present a general expression for the joint amplitude function which characterizes the photon pairs, and illustrate the resulting behavior through numerical simulations.

 In particular we show, for an SPDC source based on a periodically-poled PPLN crystal, the effect of the $\chi^{(2)}$ grating parameters
 $D$ and $z_0$ on the resulting single photon spectrum. From these results it becomes clear that both the width and, to some extent, the shape of the single photon spectrum may be controlled through these parameters. While it has been known for some time that a chirped QPM crystal leads to broadening of the SPDC spectrum, it is interesting to investigate at which times within the biphoton wavepacket specific frequencies are emitted. From our simulations of the CWF it becomes evident, for the $D\neq0$ case, that the signal-mode heralded single photon indeed exhibits a correlation between the emission frequency and the time of emission, i.e. a single-photon chirp.

 We have also shown that the degree of entanglement, as characterized by the Schmidt number, has a strong dependence on the crystal chirp.  While the specific crystal source employed here was not designed for factorability, and thus the resulting Schmidt number is far from unity, it becomes clear from our results that crystal chirp could be used as an effective tool for tailoring the degree of entanglement.  We hope that these results will be useful in the pursuit of high-flux photon pair sources, while achieving an effective control over the spectral and temporal properties of the emitted signal and idler photons.

\section*{Acknowledgments}
J.L.L.M and X.J.S.L. acknowledge financial support from CONACyT under contract CB2009-136186-F-1774. This work was supported in part by CONACYT-Mexico, by DGAPA and UNAM.

\end{document}